\documentclass{rnaastex}

\begin{document}

\title{Gaia 1 cannot be a Thick Disk Galactic cluster}

\correspondingauthor{Giovanni Carraro}
\email{giovanni.carraro@unipd.it}

\author[0000-0002-0155-9434]{Giovanni Carraro}
\affiliation{Department of Physics and Astronomy, Padova University\\
Vicolo Osservatorio 3\\
I-35122, Padova,Italy}


\section{} 

Gaia 1 was discovered by \citep{2017MNRAS.470.2702K} as a stellar over-density in the Gaia satellite first data release and originally classified as
a satellite of the Milky Way.  Following medium- and high-resolution spectroscopic investigations by \citep{2017MNRAS.471.4087S} , \citep{2017A&A...603L...7M},  and \citep{2017A&A...609A..13K} revealed that Gaia 1 is, more conservatively, an intermediate-age Galactic cluster, previously overlooked because of its proximity to the bright star Sirius. 
In spite of all these works, Gaia 1 properties are far from being settled, and worrisome discrepancies exist among the various studies.
The age ranges from 3 to 6 Gyr, which is well within the domain of Milky Way (MW) old open clusters.
Metallicity is found to go from  [Fe/H]=0  to [Fe/H]=-0.6 (virtually a factor of 10 difference), which represents almost the entire range of metallicity for MW old open clusters.
Because of the well-known age metallicity degeneracy, the isochrone fit works fine in both extremes.
It is particularly difficult to understand the reasons for such huge differences, and we look forward for a prompt solution of this puzzle, which makes any spectroscopic inference weak at present.\\

\begin{figure}[h!]
\begin{center}
\includegraphics[scale=0.85,angle=0]{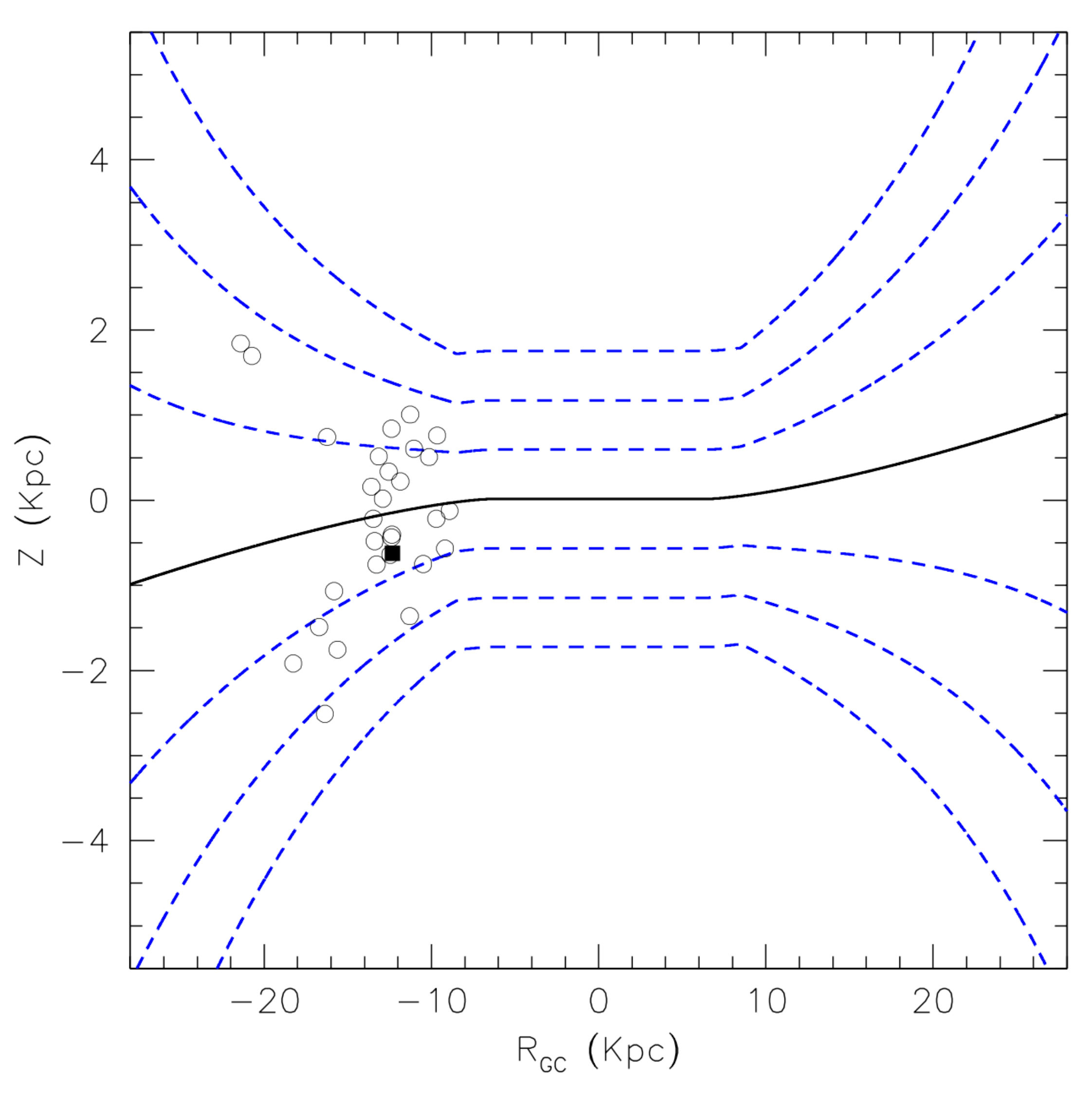}
\caption{A cut in the YZ plane of the warped and flared Galaxy. The thick line marks the mean warped stellar disk whereas the blue dashed lines trace the density at 1×, 2× and 3× the scale-height of the disk. The filled square marks the location of Gaia 1. With open circles anti-center old open clusters from \citep{2007A&A...476..217C}  are shown.\label{fig:1}}
\end{center}
\end{figure}

\noindent
Meanwhile there is an issue which has not been tackled properly by these studies, and which I want to discuss in this note.
If the metallicity is  -0.6 and the age 6  Gyr \citep{2017A&A...609A..13K}, Gaia 1 would bear a remarkable similarity with Whiting 1 \citep{2007A&A...466..181C}.
What, however, makes Whiting 1 not an open cluster is its position in the Galactic halo, as amply discussed in \citep{2007A&A...466..181C}. 
In the case of Gaia 1
\citep{2017A&A...609A..13K} conclude that Gaia 1 is a thick disk cluster because of its  actual location, almost 1 kpc above the {\it formal} (b=0$^o$) Galactic plane.
They also derived the Galactic orbit of the cluster and found contradicting results. On one side, the eccentricity 
(e= 0.12) is found be compatible with thin disk stars, but the maximum height above the plane, Zmax =1 kpc is according to them too high for the thin disk and more compatible with the thick disk.
If true, Gaia 1 would be the first open cluster ever associated with the thick disk,  which is well known to be devoid of such clusters. 
The orbit integration is done for a perfectly flat  and time independent disk potential - and run for 10 Gyr ! -  while the disk of the Milky Way is known to exhibit a 
significant warp and a significant flare \citep{2006A&A...451..515M}, which are prominent also in the direction of Gaia 1 \citep{2007A&A...476..217C}. 
This is illustrated in Figure \ref{fig:1}, where thin disk old open clusters' locations from \citep{2007A&A...476..217C} are confronted with the warped and flared Galactic disk
\citep{2006A&A...451..515M}.
Not surprisingly, Gaia 1 (the filled square) falls nicely in the distribution of such objects and is  located less than half a kpc from the {\it real}  Galactic disk. 
This results rules out its association with the Galactic thick disk, and lend additional support to the nature of Gaia 1 as  yet another thin disk old open cluster \citep{2017A&A...603L...7M}.

\acknowledgments

The author thanks Yazan Momany for his help to prepare Figure 1.

\end{document}